\documentclass[12pt]{article}\usepackage{a4}

\newcommand{\res}{{\raisebox{-6pt}{$\begin{array}{c}{\rm \normalsize Res}\\ s=\frac32-n\end{array}$}}}
\newcommand{\be}{\begin{equation}}
\newcommand{\ee}{\end{equation}}
\newcommand{\bea}{\begin{eqnarray}}
\newcommand{\eea}{\end{eqnarray}}
\newcommand{\beq}{\begin{eqnarray}}
\newcommand{\eeq}{\end{eqnarray}}
\newcommand{\beao}{\begin{eqnarray*}}
\newcommand{\eeao}{\end{eqnarray*}}
\newcommand{\Ref}[1]{(\ref{#1})}

\newcommand{\hkks}{heat kernel coefficients }
\newcommand{\hkksp}{heat kernel coefficients. }
\newcommand{\hkk}{heat kernel coefficient }

\newcommand{\nn}{\nonumber}
\newcommand{\pa}{\partial}

\newcommand{\ep}{{\epsilon}}
\newcommand{\om}{{\omega}}
\newcommand{\gse}{ground-state energy }
\newcommand{\uv}{ultraviolet }

\begin{document}

\title{Heat Kernel Coefficients and Divergencies of the Casimir Energy
for the Dispersive Sphere}
\author{{M. Bordag}\thanks{e-mail: Michael.Bordag@itp.uni-leipzig.de} \\
\small  University of Leipzig, Institute for Theoretical Physics\\
\small  Augustusplatz 10/11, 04109 Leipzig, Germany\\[.2cm]
{K. Kirsten}\thanks{e-mail: kirsten@mis.mpg.de}\\
\small Max-Planck-Institute for Mathematics in the Sciences\\
\small Inselstr. 22-26, 04103 Leipzig, Germany}
\maketitle
\begin{abstract}
  The first heat kernel coefficients are calculated for a dispersive ball
  whose permittivity at high frequency differs from unity by inverse powers
  of the frequency. The corresponding divergent part of the vacuum energy of
  the electromagnetic field is given and ultraviolet divergencies are seen to
  be present. Also in a model where the number of atoms is fixed the
  pressure exhibits infinities. As a consequence, the ground-state energy 
  for a dispersive dielectric ball cannot be interpreted easily.
\end{abstract}

The \gse for a dielectric ball shows ultraviolet divergencies still
lacking physical understanding.  This is an unsatisfactory situation,
not only for general reasons but also in view of the rapid
experimental progress.

The canonical way to investigate the \uv divergencies is to calculate
the corresponding \hkksp For the dielectric (nondispersive) ball this
had been done in \cite{Bordag:1999vs} and for the dielectric cylinder
in \cite{Bordag:2001zj}, where it had been shown, for instance, that the
coefficient $a_2$ is zero in dilute order and nonzero beyond.  In the
present note we calculate the relevant \hkks for a dielectric ball
with dispersion.

Dispersion means a frequency dependent permittivity, $\ep(\om)$. This
is motivated by the expectation that for high frequency the
permittivity tends to unity 
so that the ultraviolet modes  contribute less to the \gse and
the divergencies become weaker. 

It is reasonable to assume the
asymptotic behavior of $\ep(\om)$ to be
\be\label{epom} \ep(\om)=1-\frac{\ep_1}{\om^2}+\frac{\ep_2}{\om^4}+\dots
\ee
for $\om\to\infty$. Higher terms do not contribute to the \uv divergencies.
Let us note that this asymptotic behavior is typical for solid state models,
the Drude and plasma models for instance. In the latter, the parameter $\ep_1$
in Eq. \Ref{epom} has the meaning of the plasma frequency squared.

We remind the reader of 
some basic formulas. In zeta functional regularization the
\gse is given by \cite{eliz94b,eliz95b}
\be\label{gse} E_0(s)=\frac{\mu^2}{2}\sum_j\lambda_j^{1-2s},
\ee
where $\lambda_j$ 
are the corresponding (discrete) energy eigenvalues. It can be
expressed in terms of the corresponding zeta function,
\be\label{e0s}E_0(s)=\frac{\mu^2}{2}\zeta(s-\frac12).
\ee
Here $\mu$ is an arbitrary parameter with the dimension of a mass. Dropping
the so called Minkowski space contribution the zeta function can be
represented as
\be\label{zeta}\zeta(s)=\frac{\sin \pi s}{\pi} \sum_{l=1}^\infty(2l+1)
\int_0^\infty dk \ k^{-2s} \frac{\pa}{\pa k} \ln f_l(ik),
\ee
where $ f_l(ik)$ is the Jost function of the corresponding scattering
problem. A detailed explanation of these and related formulas can be
found in \cite{Bordag:1999vs,Bordag:1996fv,Kirsten:2000xc}.

The \hkks can be obtained from the zeta function by means of 
\be\label{hkks} a_n=(4\pi)^{3/2}\res \Gamma(s)\zeta(s)
\ee
and the divergent part of the \gse in zeta functional regularization is
given by (for a massless field)
\be\label{disgse}E_0^{\rm div}(s)=\frac{-a_2}{32\pi^2}\left(\frac{1}{s}+2\ln \mu-2\right).  \ee
Here we drop contributions from $a_{1/2}$ and $a_{3/2}$, as these coefficients
will turn out to vanish for the problem considered.

It is known that in the zeta function regularization there is a smaller number
of singular contributions to the vacuum energy than in other regularization
schemes.  For example, in the regularization
\be\label{e0d}E_0(\delta)=\frac12 \sum_j \lambda_j \ e^{-\delta \lambda_j}
\ee
with an exponentially damping function the divergent part of the \gse is 
\be\label{e0ddiv} 
E_0^{\rm div}(\delta)=\frac{1}{16\pi^2}\left(\frac{2}{\delta^2}a_1+\ln\delta \
  a_2\right),
\ee
again with the Minkowski space contribution already subtracted and 
dropping the terms with
$a_{\frac12}$ and $a_{\frac32}$.
 
The Jost function for the problem at hand is known. 
For convenience, during the calculation we put the radius $R$ of the 
dielectric ball equal to one, $R=1$. The dependence on $R$ is recovered 
by dimensional arguments. With the notation $\nu =  l+1/2$, 
the Jost function consists of
contributions from the TE and the TM modes,
\be\label{f} f_l(ik)=\Delta^{TE}_\nu(ik) \ \Delta^{TM}_\nu(ik)
\ee
($l=1,2,\dots)$ with
\bea\label{TE}
\Delta^{TE}_\nu(ik)&=&\ep^{-\frac{\nu}{2}}\left(\sqrt{\ep}s'e-se'\right),\\
\label{TM}
\Delta^{TM}_\nu(ik)&=&\ep^{-\frac{\nu}{2}}
\left(\frac{1}{\sqrt{\ep}}s'e-se'\right).  \eea
Here the abbreviations 
\bea  s\equiv s_l(q)&=&\sqrt{\frac{\pi q}{2}} I_\nu(q), \\
      e\equiv e_l(k)&=&\sqrt{\frac{2k}{\pi}} K_\nu(k),
\eea
are used where $I_\nu(q)$ and $K_\nu(k)$ are the modified Bessel
functions. The prime denotes the differentiation with repect to the
argument of these functions. The arguments of the Bessel functions
are related by
\be\label{ep} q=\sqrt{\ep} k.
\ee
The factors $\ep^{-\frac{\nu}{2}}$ in \Ref{TE} and \Ref{TM} follow
from the normalization condition of the regular solution of the
scattering problem, for details see the example of a square well
potential in \cite{Bordag:1996fv}. This is of importance since we
consider $\ep$ depending on $k$. We mention that these representations hold
in the presence of arbitrary frequency dispersion, as has
been noted in \cite{brev00-33-5819} (see also \cite{brev94-49-5319}).

In order to get the residues according to Eq. \Ref{hkks} it is
sufficient to approximate the Jost function by its 
uniform asymptotic expansion 
for large $k$ and $\nu$ keeping $z\equiv \frac{k}{\nu}$
fixed.  Using the well known expansions \cite{abra70b}, we obtain
\bea\label{1exp} \ln
f_l(ik)&=&\nu\left(2\left(\eta(\sqrt{\ep}z)-\eta(z)\right)-\ln\ep\right)
 \nn\\
&&\hspace{2cm}+\ln\left(\ep^{ \frac14} \tilde{s}'\tilde{e}
-\ep^{-\frac14}\tilde{s}\tilde{e}'\right)+\ln\left(\ep^{-\frac14}
\tilde{s}'\tilde{e}-\ep^{ \frac14}\tilde{s}\tilde{e}'\right) \nn \\
&\equiv&D_0+D_{TE}+D_{TM}, \eea
where the tilde denotes the Bessel functions without the exponential factors. 
Now we use the expansion
\be\label{epik} \ep(ik)=1+\frac{\ep_1}{k^2}+\frac{\ep_2}{k^4}+\dots
\ee
as well as the known expression for $\eta$, for example
$\eta'(z)=\sqrt{1+z^2}/z$, and obtain
\be\label{D0}
D_0=\frac1\nu \ep_1\frac{\sqrt{1+z^2}-1}{z^2} 
+\frac{1}{\nu^3} \left(\ep_2\frac{\sqrt{1+z^2}-1}{z^4}
-\frac{\ep_1^2}{4} \ \frac{(\sqrt{1+z^2}-1)^2 }{z^4\sqrt{1+z^2}}\right)+\dots
\ee
For $D_{TE}$ and $D_{TM}$ we obtain
\be\label{DTE} D_{TE}=\ln \left\{\frac12 
\left[ 
\left(\frac{1+\ep z^2}{1+z^2}\right)^{\frac14}\left(1+C\right)\left(1+B\right) 
+
\left(\frac{1+\ep z^2}{1+z^2}\right)^{-\frac14}\left(1+A\right)\left(1+D\right)
\right]\right\}
\ee
and  
\bea\label{DTM} D_{TM}&=&\ln \left\{\frac12 \left[
\ep^{-\frac12}\left(\frac{1+\ep
z^2}{1+z^2}\right)^{\frac14}\left(1+C\right)\left(1+B\right) +
\ep^{\frac12}\left(\frac{1+\ep
z^2}{1+z^2}\right)^{-\frac14}\left(1+A\right)\left(1+D\right) \right]
\nn \right. \\ && \left.
+\left(\ep^{-\frac12}-\ep^{\frac12}\right)\frac{(1+A)(1+B)}{4\nu((1+z^2)(1+\ep
z^2))^\frac14 } \right\} , \eea
where we used the same abbreviations for the Debye polynomials as in
\cite{Bordag:1999vs}. We need them in the first nontrivial order only,
$A=u_1(t_q)/\nu$, $B=-u_1(t)/\nu$, $C=v_1(t_q)/\nu$ and
$D=-v_1(t)/\nu$ with $t=1/\sqrt{1+z^2}$ and $t_q=1/\sqrt{1+\ep
z^2}$. Inserting now the expansion \Ref{epik} of $\ep$ we obtain
finally
\be\label{DTEf}D_{TE}=-\frac{1}{\nu^3}
\frac{\ep_1}{16}\frac{z^2}{(1+z^2)^{5/2}}+\dots \ee
and
\be\label{DTMf} D_{TM}=-\frac{1}{\nu^3}
\frac{\ep_1}{16}\frac{z^4+4z^2+4}{z^2(1+z^2)^{5/2}}+\dots \ . \ee
We have to insert these expansions, Eqs. \Ref{D0}, \Ref{DTEf},
\Ref{DTMf}, into the Jost function, Eq. \Ref{f}, and the latter, then,
into the zeta function, Eq. \Ref{zeta}. Performing there the variable
substitution $k=\nu z$, the sum over $\nu$ can be carried out
which gives Hurwitz zeta functions,
\be\label{zetaH}
\zeta_H(s;\frac32)=\sum_{l=1}^\infty\left(l+\frac12\right)^{-s}.  \ee
In summary, we obtain
\bea\label{zetaint} \zeta(s)&=&2\frac{\sin \pi s}{\pi} \left\{
\zeta_H(2s;\frac32)\int_0^\infty dz \ z^{-2s}\frac{\pa}{\pa z}
\ep_1\frac{\sqrt{1+z^2}-1}{z^2} \right.\\ && +
\zeta_H(2s+2;\frac32)\int_0^\infty dz \ z^{-2s}\frac{\pa}{\pa z}
\left[ \ep_2 \frac{\sqrt{1+z^2}-1}{z^4}
-\frac{\ep_1^2}{4}\frac{(\sqrt{1+z^2}-1)^2}{z^4 \sqrt{1+z^2}}
\right.\nn \\ && \left.\left. -\frac{\ep_1}{8}\frac{z^4+2z^2+2}{z^2
(1+z^2)^{5/2}}\right]\right\} \ . \nn \eea

Now it is easy to extract the \hkks using Eq. \Ref{hkks}. The
rightmost pole is at $s=1/2$ resulting from the first Hurwitz zeta
function. It yields
\be\label{a1}  a_1=-\frac{8\pi}{3}\ep_1.
\ee
The next pole is at 
$s=-\frac12$. It results from the integral in the first line of 
Eq. (\ref{zetaint}), 
\[ 
\int_o^\infty dz \ z^{-2s}\frac{\pa}{\pa
z}\frac{\sqrt{1+z^2}-1}{z^2}=\frac{-1}{2\sqrt{\pi}}
s\Gamma(-1-s)\Gamma(s+\frac12).
\]
Further contributions result from the pole of the second Hurwitz zeta
function. Taking all contributions together we obtain
\be\label{a2} a_2=\frac{4\pi}{3}\ep_1^2+\frac{16\pi}{3}\ep_2 ,  \ee
where a term linear in $\ep_1$ cancelled out between TE and TM
contributions. To $a_1$, \Ref{a1}, and $a_2$, \Ref{a2}, the TE and TM
modes give equal contributions.

Formulas \Ref{a1} and \Ref{a2} are the main result of this paper. 
In order to draw more physical conclusions we restore the 
dependence on $R$. From
$[\ep_{1}]=R^{-2}$, $[\ep_{2}]=R^{-4}$, $[a_1]=R$ and $[a_2]=R^{-1}$ we get
\[a_1=-\frac{8\pi}{3}\ep_1R^3, \qquad a_2=\frac{4\pi}{3}\ep_1^2 R^3 
+\frac{16\pi}{3}\ep_2 R^3 .
\]
Now the divergent part of the \gse reads in zeta functional regularization
\be\label{e0sdiv1} 
E_0^{\rm div}(s)=\frac{-1}{32\pi^2}\left(\frac1s+2\ln\mu-2\right)
\left(\ep_1^2+4\ep_2\right)V, \ee
where $V=\frac{4\pi}{3}R^3$ is the volume of the ball, and in the
regularization Eq. \Ref{e0d}, using the exponentially damping function, 
it is
\be\label{e0ddiv1}
E_0^{\rm
  div}(\delta)=\frac{1}{16\pi^2}\left(\frac{-2\ep_1}{\delta^2}+\left(\ep_1^2+4\ep_2\right)\ln\delta\right)V.
\ee
These results 
show that for a dispersive dielectric ball \uv divergencies are present
in a fashion similar to that for the nondispersive case.\\

\noindent We add some remarks.
\begin{enumerate}
\item In both regularizations the \gse energy is divergent. Because of
  $a_2\ne0$, even after the removal of the diverging contributions an
  arbitrariness (e.g., $\ln\mu$ in Eq. \Ref{e0sdiv1}) remains in the finite
  part.
  
  The model, Eq. \Ref{epom}, chosen for the permittivity reflects the physical
  assumption that for high frequencies the dielectric ball becomes
  transparent.  From the results, Eqs. \Ref{e0sdiv1} and \Ref{e0ddiv1},
  in particular from the contribution of $\ep_2$, it follows that this is
  insufficient in order to get a satisfactory physical interpretation. 
  Although the dependence of the divergencies on the dielectric properties
  is considerably simpler than in \cite{Bordag:1999vs}, we
  are left with the same 
  conclusions as for the
  nondispersive case that for a dielectric body substantial changes in the
  physical context are necessary \cite{Bordag:1999vs}.
\item In the sense of renormalization one might try to absorb the divergent
  contributions into some classical part.  For example, in the bag model, \uv
  divergent contributions proportional to the volume like Eqs. \Ref{e0sdiv1}
  and \Ref{e0ddiv1} can be put into a redefinition of the bag
  constant. However, in the present case we do not have any classical energy
  which could be associated with the dielectric ball. In addition, we don't
  have any normalization condition which might help to fix the arbitrariness. 
\item As compared with the nondispersive case, \cite{Bordag:1999vs}, the non
  vanishing of $a_2$ is a common feature. The only known exception is the
  vanishing of $a_2$ in the dilute approximation, i.e., to order $(\ep-1)^2$
  for $\ep\to1$, which allowed the \gse to have a physical meaning and ensures
  that the results of different calculations coincide. However, as it was
  shown in \cite{Bordag:1997fh} and \cite{Bordag:1999vs}, this is a 
  peculiarity
  of a ball with sharp boundaries.  For a dielectric body with non sharp
  boundaries, i.e., with the permittivity $\ep(r)$ being a smooth function
  of the radius, $a_2$ is non zero even in the dilute approximation.
\item One might hope that the pressure (force per unit surface) is \uv finite
  rather than the vacuum energy itself. For this end one has to divide by the
  surface ($4\pi R^2$) of the ball and to take the derivative with respect to
  the radius.  As the divergent part of the \gse is proportional to the volume
  of the ball the pressure contains a divergent constant.
\item One may assume the dielectric ball to be an idealization of a number of
  polarizable atoms. Then a change in the radius leaving the number of atoms
  fixed requires a change in $\ep$ according to $(\ep-1)V=const$ as discussed,
  for instance, 
  in \cite{Barton:2001,Marachevsky:2001pc,Marachevsky:2000yi}. In
  this case, by means of Eq. \Ref{epom}, $\ep_{1,2}V=const$ follows. Due to
  the presence of $\ep_1^2$ in \Ref{e0sdiv1} and \Ref{e0ddiv1}, again, a
  divergent contribution is present.
\item An investigation similar to the present one
  had been recently carried out in
  \cite{Falomir:2001uv}, where the divergent part of the Casimir energy had
  been calculated for the plasma model in zeta functional regularization. This
  is equivalent to calculate the contribution of $\ep_1$ to the \hkk $a_2$ and
  is in agreement with Eq. \Ref{a2}.
\item In the perturbative approach described in \cite{Barton:2001} 
  divergencies proportional to the surface area were found. It would
  be very desirable to  understand the origin of the different 
  predictions in the different schemes used.
\end{enumerate}
\section*{Acknowledgments}
We thank D. Vassilevich for useful discussions. KK is grateful for the 
support provided by the MPI for Mathematics in the Sciences, Leipzig.

\end{document}